\documentclass[aps,prl,reprint,10pt,superscriptaddress]{revtex4-2}
\usepackage[left=2.5cm,right=2.5cm, top=3cm,bottom=3cm]{geometry}
\usepackage[english]{babel}
\usepackage{amsmath,amssymb,amsthm,mathtools,bm}
\usepackage{euscript,mathrsfs}
\usepackage{graphicx}
\usepackage{wrapfig}
\usepackage[dvipsnames]{xcolor}
\usepackage{indentfirst}
\usepackage{amsbsy}
\usepackage{cancel}
\usepackage{verbatim}
\usepackage{empheq}
\usepackage{adjustbox}
\usepackage[pdfencoding=auto]{hyperref}
\definecolor{urlcolor}{HTML}{990000}
\definecolor{linkcolor}{HTML}{005F5F}
\hypersetup{pdfstartview=FitH,  linkcolor=linkcolor,urlcolor=urlcolor, colorlinks=true,citecolor=blue}

\renewcommand{\phi}{\varphi}
\renewcommand{\epsilon}{\varepsilon}

\DeclareMathOperator{\1}{\mathbf{1}}

\newcommand{\iu}{\mathrm{i}}

\newcommand{\R}{\mathbb{R}}
\newcommand{\T}{\mathbb{T}}

\newcommand{\Z}{\mathbb{Z}}
\newcommand{\N}{\mathbb{N}}
\newcommand{\Prob}{\mathbb{P}}

\newcommand{\bW}{\boldsymbol{W}}
\newcommand{\bK}{\boldsymbol{K}}

\newcommand{\cF}{\mathcal{F}}
\newcommand{\cL}{\mathcal{L}}

\usepackage[dvipsnames]{xcolor}
\usepackage{tikz-cd}
\usepackage{tikz}
\usetikzlibrary{positioning}
\usetikzlibrary{calc}

\definecolor{mylightred}{RGB}{211,79,73}
\definecolor{mydarkred}{RGB}{199,44,38}
\definecolor{mylightgreen}{RGB}{78,153,67}
\definecolor{mydarkgreen}{RGB}{43,129,33}
\definecolor{mylightpurple}{RGB}{150,107,178}
\definecolor{mydarkpurple}{RGB}{126,78,160}
\definecolor{mylightblue}{RGB}{49,101,205}
\definecolor{mydarkblue}{RGB}{20,92,205}

\tikzset{
  juliadot/.style args={#1,#2}{shape=circle,line width=0.03ex,minimum width=0.4ex,fill=#1,draw=#2}
}

\begin{document}

\title{Phase transitions in the Ising model on random graphs}
\author{Artem Alexandrov}
\email{aleksandrov.aa@phystech.edu}
\affiliation{HSE University, Moscow, Russia}
\affiliation{Phystech School of Applied Mathematics and Computer Science, Moscow Institute of Physics and
  Technology, Dolgoprudny 141700, Russia}
\affiliation{Laboratory of Complex Networks, Center for Neurophysics and Neuromorphic Technologies,
  Moscow, Russia}

\author{Georgi Medvedev}
\email{medvedev@drexel.edu}
\affiliation{Department of Mathematics, Drexel University, 3141 Chestnut Street, Philadelphia, PA 19104}

\begin{abstract}
  We study phase transitions in the Ising model on random graphs using graph limits. This framework
  extends mean-field theory to heterogeneous nonlocal interactions. We show that the critical
  temperatures are determined by the eigenvalues of the kernel operator associated with the graph
  limit. Bifurcation diagrams for Erdős–Rényi, small-world, and power-law graphs illustrate the theory.
  In the small-world case,
  we identify metastable behavior in both ferromagnetic and antiferromagnetic regimes.
\end{abstract}

\maketitle

The Ising model with nearest-neighbor interactions provides a theoretical framework for
understanding magnetism. Exactly solvable mean-field models have been
used to gain insights into the mechanisms of phase transitions in the Ising model \cite{BovKur2009}.
The graphon Ising model, recently introduced by Searle and Tindall \cite{SeaTin2024},
combines the analytical tractability of mean-field models with the flexibility to incorporate
diverse connectivity patterns. Using techniques from graphon dynamical
systems \cite{Med2014a,Med19,MedPel2024}, we analyze phase transitions in the Ising model
on random graphs, highlighting the roles of connectivity and symmetry. We illustrate our results
through detailed analyses of the
Ising model on Erd\H{o}s–Rényi, power-law, and small-world graphs.

Consider the Ising model on graph $\Gamma_n$ on $n$ model labeled by
$[n]\doteq \{1,2,\dots, n\}$ with adjacency matrix $A=(a_{ij})$.
At each node~$i\in[n]$, we place $\nu$ atoms with spins $\sigma_i^s\in\{-1, 1\}, s\in [\nu]$.
The state of the system is determined by the Hamiltonian
\begin{equation}\label{pre-Ham}
  H(\sigma)=-\frac{1}{2 n\nu^2}\sum_{i,j=1}^n\sum_{s,t=1}^\nu J_{ij} \sigma_i^s\sigma_j^t
  -h\sum_{i,s=1}^{n,\nu}\sigma_i^s,
\end{equation}
where $J_{ij}$ is the strength of pairwise interactions $h$ is external magnetic field,
which will be assumed to be $0$ throughout this Letter.

Upon averaging spins at each node, we obtain local magnetization 
$
m_i\approx \nu^{-1}\sum_{s=1}^\nu\sigma_i^s.
$
The free energy is then expressed as
$$
F(m)=\frac{-1}{2n}\sum_{i,j=1}^n J_{ij} m_im_j -T\sum_{i=1}^nS(m_i),
$$
where $S(x)=-\left(\ell(\frac{1+x}{2})+\ell(\frac{1-x}{2})\right),$ $\ell(x)=x\log x$.
After setting $\partial_{m_i}F(m)=0$, we arrive at
\begin{equation}\label{steady-i}
  m_i=\tanh\left( \beta n^{-1}\sum_{j=1}^n a_{ij} m_j\right), \; i\in [n],
\end{equation}  
where we used $J_{ij}=Ja_{ij}$ and $\beta=JT^{-1}$.

To ensure that the thermodynamic limit in Eq.~\eqref{steady-i} exists, the graph sequence $\Gamma_n$
must converge in an appropriate sense. Dense graph convergence \cite{LovSze2006}, and more generally
$L^p$-graphon convergence for sparse graph sequences \cite{BCCZ19}, provide the appropriate
mathematical framework.

Next, we specify the $W$-random graph model, which will serve as
our primary example of a convergent graph sequence. Let $W: Q \times Q \to [0,1]$ be a symmetric
measurable function (throughout this Letter, $Q\doteq [0,1]$). In the theory of graph limits,
such functions are called graphons \cite{Lovasz-book}. Partition $Q$ into $n$ subintervals
$Q_i^n = [(i-1)/n, i/n)$ and average $W$ over $Q_{ij}^n = Q_i^n \times Q_j^n$:
$
W^n_{ij}=n^2\int_{Q^n_{ij}} W(x,y)dxdy, \; i,j\in [n].
$

The entries of the adjacency matrix $a_{ij}, 1\le i\le j\le n,$ are defined as independent binary
random variables with
$\Prob(a_{ij}=1)=W^n_{ij}$ and $\Prob(a_{ij}=0)=1-W^n_{ij}$. To derive the continuum limit for
\eqref{steady-i}, we follow
the scheme developed in \cite{Med19}. First, rewrite \eqref{steady-i} as an integral equation
$$
m^n(x)=\tanh\left(\beta\int_Q A^n(x,y)m^n(y)dy\right), \; x\in Q,
$$
where $A^n=\sum_{i,j=1}^n a_{ij} \1_{Q^n_i}(x)\1_{Q^n_j}(y)$ and $m^n=\sum_{i=1} m_i\1_{Q^n_i}(x),$ and
$\1$ stands for the indicator function. Since $A^n$ converges to $W$ in the cut-norm almost surely
(cf.~\cite{DupMed2022}), the natural candidate for the thermodynamic limit of \eqref{steady-i} is
\begin{equation}\label{clim}
\cF(u,\beta)\doteq\tanh\left(\beta\bW[u] \right) -u(x)=0,
\end{equation}
where $\bW[u]\doteq \int_Q W(\cdot,y)u(y)~dy$.


The trivial solution $u\equiv 0$ (a paramagnetic phase) exists for all $\beta$.
The phase transitions in the Ising model \eqref{pre-Ham} correspond to the bifurcations
of the trivial solution of \eqref{clim}. The latter are determined by the spectrum of
the kernel operator
$\bW$
and the symmetries in \eqref{clim}. As a self-adjoint Hilbert-Schmidt operator on $L^2(Q)$,
$\bW$ has at most a countable number of eigenvalues with a single
accumulation point at $0$:
\begin{equation}\label{spec-W}
 \lambda_1\ge \lambda_2\ge \dots\ge \lambda_0=0\ge \dots\ge
\lambda_{-2}\ge \lambda_{-1}.
\end{equation}
Denote the spectrum of $\bW$, by $\sigma(\bW)$.

Let $\beta=\beta_0+\gamma$ and
rewrite \eqref{clim} as follows
\begin{equation}\label{set-F}
\cL_{\beta_0}[u]+\gamma \bW[u]
-\frac{\beta_0^3}{3} \left(\bW[u]\right)^3+\dots=0,
\end{equation}
where 
$\cL_{\beta_0}[u]\doteq \beta_0\bW[u]-u$.

Note that if 
$
\beta_0^{-1} \notin \sigma(\bW),
$ 
then $\cL_{\beta_0}$ has a bounded inverse,
and by the Implicit Function Theorem, 
$
u \equiv 0
$ 
is the unique solution of \eqref{clim} for small $|\gamma|$ \cite{Dieu-Analysis}.

If
$
\beta_0^{-1} = \lambda_k \in \sigma(\bW),
$ 
a phase transition is expected. The odd symmetry 
$
\cF(-u,\beta) = -\cF(u,\beta)
$ 
indicates a pitchfork bifurcation.

Thus, the eigenvalues of the kernel operator determine to the 
critical values of the temperature at which phase transitions occur.
If the operator $\bW$ has distinct nonzero eigenvalues, multiple bifurcations may occur,
leading to the coexistence of solutions of Eq.~\eqref{clim} and hence to \textit{metastability}.
Even when $W$ is a nonnegative graphon, $\bW$ can
have negative eigenvalues, giving rise to an \textit{antiferromagnetic} (AFM) phase.

We now examine the finer structure of these phase transitions for several
representative examples of $W$.
First, consider $W \equiv p \in (0,1]$, which corresponds to an Erd\H{o}s-R\'{e}nyi graph if $p<1$,
and to the complete graph if $p=1$. In this case, 
$
\lambda_1 = p
$ 
is the only nonzero eigenvalue. It is a simple eigenvalue, with corresponding eigenfunction 
$
\xi_1(x) \equiv 1.
$  

\begin{figure}[h]
    \centering
     \includegraphics[width=\linewidth]{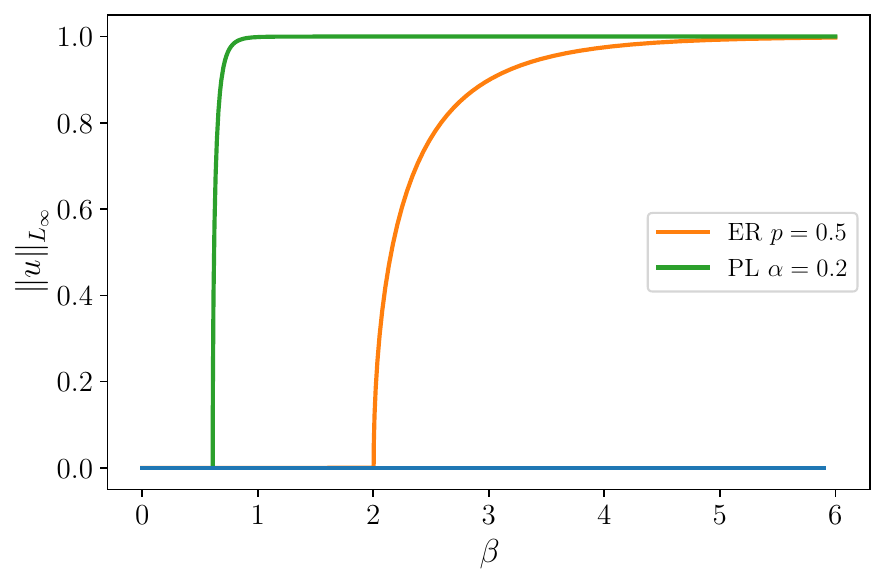}
    \caption{Bifurcations for Erd\H{o}s-R\'{e}nyi (ER) graphon with $p=0.5$ and power-law (PL) graphon
      with $\alpha=0.2$}
    \label{f.ER}
\end{figure}

By the Crandall-Rabinowitz theorem, combined with the odd symmetry of 
\(\cF(\cdot,\beta)\), we conclude that at 
$
\beta = p^{-1}
$ 
a constant (nonzero) solution emerges through a pitchfork bifurcation
(see Fig.~\ref{f.ER}).

We also consider the power-law graphon $W=(xy)^{-\alpha}$, $\alpha \in (0,1/2)$ \cite{MedTan2018}.
In this case, the operator $\bW$ has a unique nonzero eigenvalue $\lambda = (1-2\alpha)^{-1}$ of multiplicity
$1$. The bifurcation diagram in Fig.~\ref{f.ER} shows a branch of constant nonzero solutions
emerging at $\beta_c = 1 - 2\alpha$, in agreement with our theoretical prediction.

Next, we turn to our main example $ W(x,y) = K(x-y)$:
\begin{equation}\label{sw-kernel}
 K(x)=
  \begin{cases}
    1-p, & |x| \le r, \\
    p, & r < |x| \le 1/2,
  \end{cases}
\end{equation}
where $K$ is a symmetric function defined on $[-1/2,1/2]$ and extended to $\R$ by periodicity.  
For $p \in (0,1/2)$, this choice of $W$ yields $W$-random small-world graph \cite{Med2014c}.
The case $p=0$ corresponds to the nonlocal nearest-neighbor graph, where $r \in (0,1/2)$ specifies
the size of the ``nearest'' neighborhood.  
In particular, for small $r$, we obtain a model with isotropic short-range interactions.

Given the periodicity of $K$, it is natural to consider \eqref{clim}
on the torus $\T \doteq \R / \Z$ instead of the unit interval $Q$.
In this setting, $\bW$ becomes a convolution operator $\bK[u]=K\ast u$
and the eigenvalues of $\bK$ are obtained by taking the
Fourier transform of $K$:
\begin{equation}\label{eig-K}
  \mu_k=\widehat{(K)}_k\doteq\int_\T K(x)e^{-\iu 2\pi kx}dx,\; k\in\Z.
\end{equation}

\begin{figure}[h]
\begin{center}
  \includegraphics[width=\linewidth]{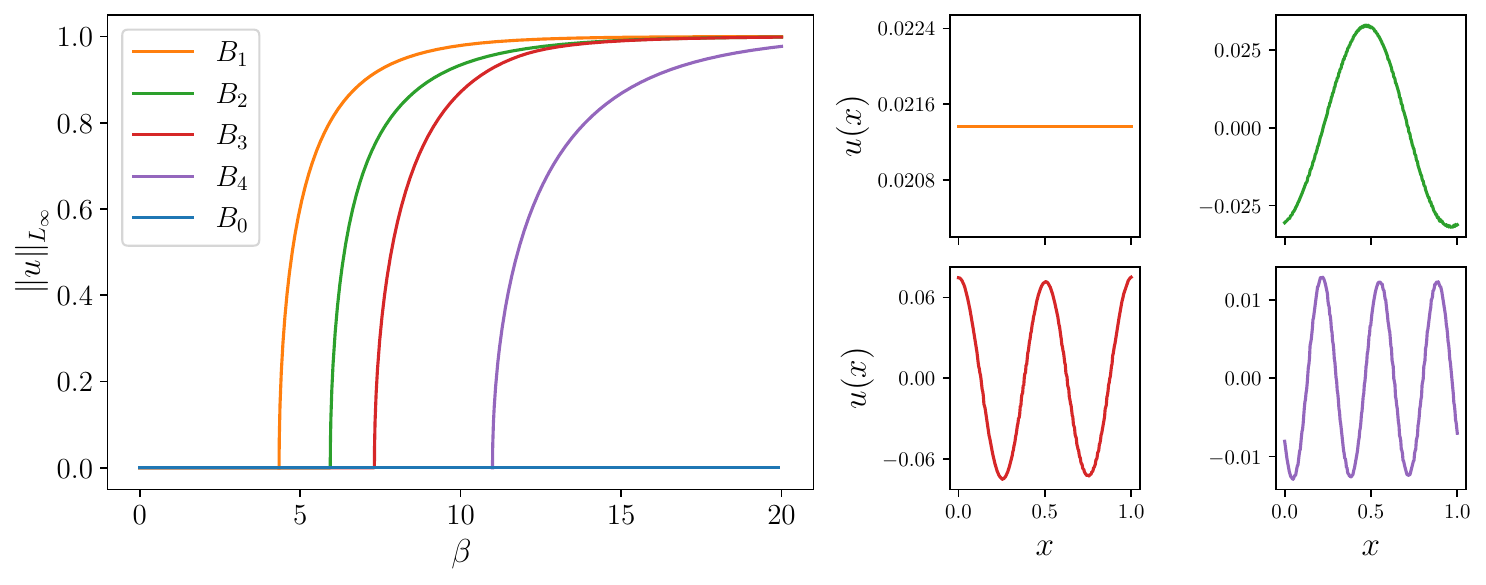}
   \includegraphics[width=\linewidth]{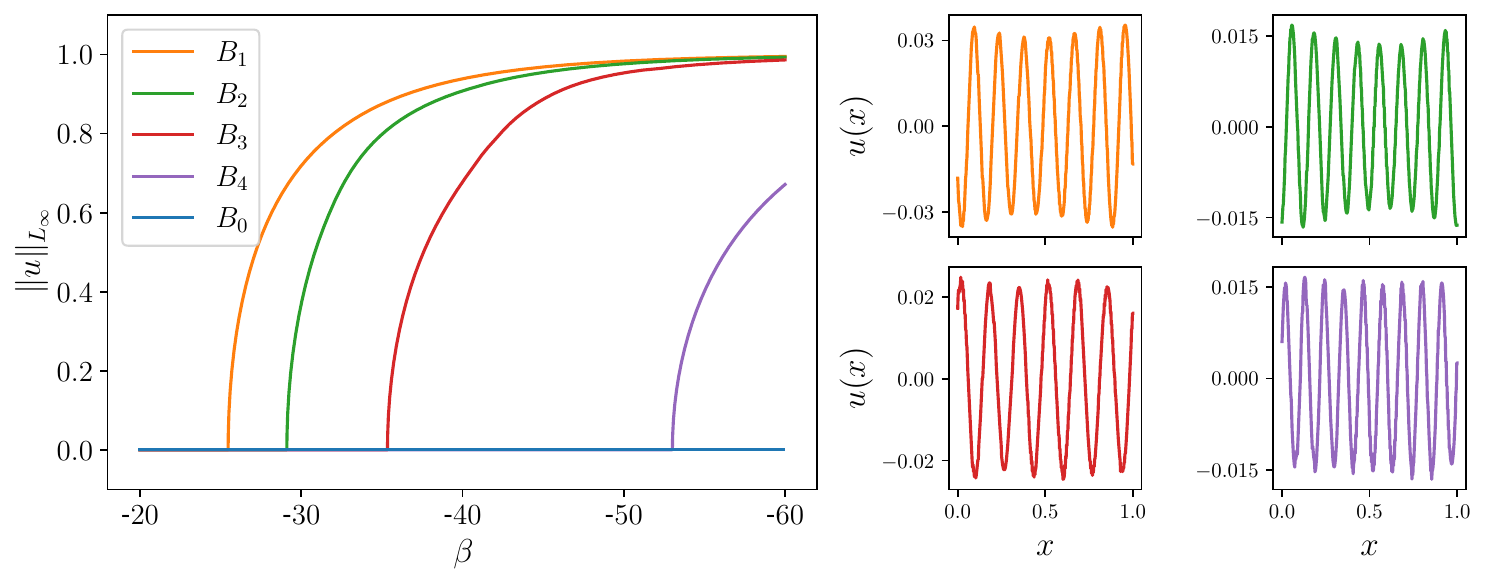}
\end{center}
\caption{Bifurcation diagrams for \eqref{clim} for the small-world
  network with $p=0.05$ and $r=0.1$. }
\label{f.bif-sw}
\end{figure}

The corresponding eigenfunctions are $\xi_k=e^{-\iu 2\pi kx}$, $k\in\Z$.
The bifurcation analysis of \eqref{clim} takes into account the following symmetries of \eqref{clim}:
  if $u(x)$ is a solution of \eqref{clim} then so are
  \begin{equation}\label{symmetries}
    u(x+h), \quad -u(x),\quad\mbox{and}\quad u(-x).
    \end{equation}
    The first follows from translation invariance of \eqref{clim}, the second from the odd symmetry
    $\cF(-u,\beta)=-\cF(u,\beta)$, and the third from the evenness
of the kernel $K(-x)=K(x)$. These symmetries determine the normal form of the
equation near the bifurcation. The pitchfork bifurcation subject to these symmetries is analyzed in detail in
\cite[\S~3]{MedPel2024}.
Here, we outline the key steps.

First, note that since $K(x)$ is even and real, $\mu_{k}=\mu_{-k}$ for all $k\in\N$.
Therefore, the multiplicity of $\mu_k$ is at least two. $\mu_0$ is simple and the corresponding
eigenfunction is constant $\xi_0\equiv 1$. Let $\beta_0=\mu^{-1}_p, p\in\N,$ and suppose the multiplicity
of $\mu_p$ is $2$. By applying the Fourier transform to \eqref{set-F}, we have
\begin{align} \nonumber 
  &\left[\frac{\mu_k}{\mu_p}\hat{u}_k - \hat{u}_k\right]  -\gamma \mu_k\hat{u}_k 
  \label{F-eqn}\\\nonumber
  &+
    \frac{1}{3\mu_p^3} 
    \sum_{k_1,k_2\in\Z}\mu_{k_1}\mu_{k_2}\mu_{k-k_1-k_2} \hat{u}_{k_1}\hat{u}_{k_2}\hat{u}_{k-k_1-k_2}+\dots\\
&=0,
\end{align}
for $k\in\Z$. Here, we used $\widehat{K\ast u}_k=\mu_k\hat{u}_k$, and $\beta_0=\mu_p^{-1}$.
\begin{figure}
\begin{center}
  \includegraphics[width=\linewidth]{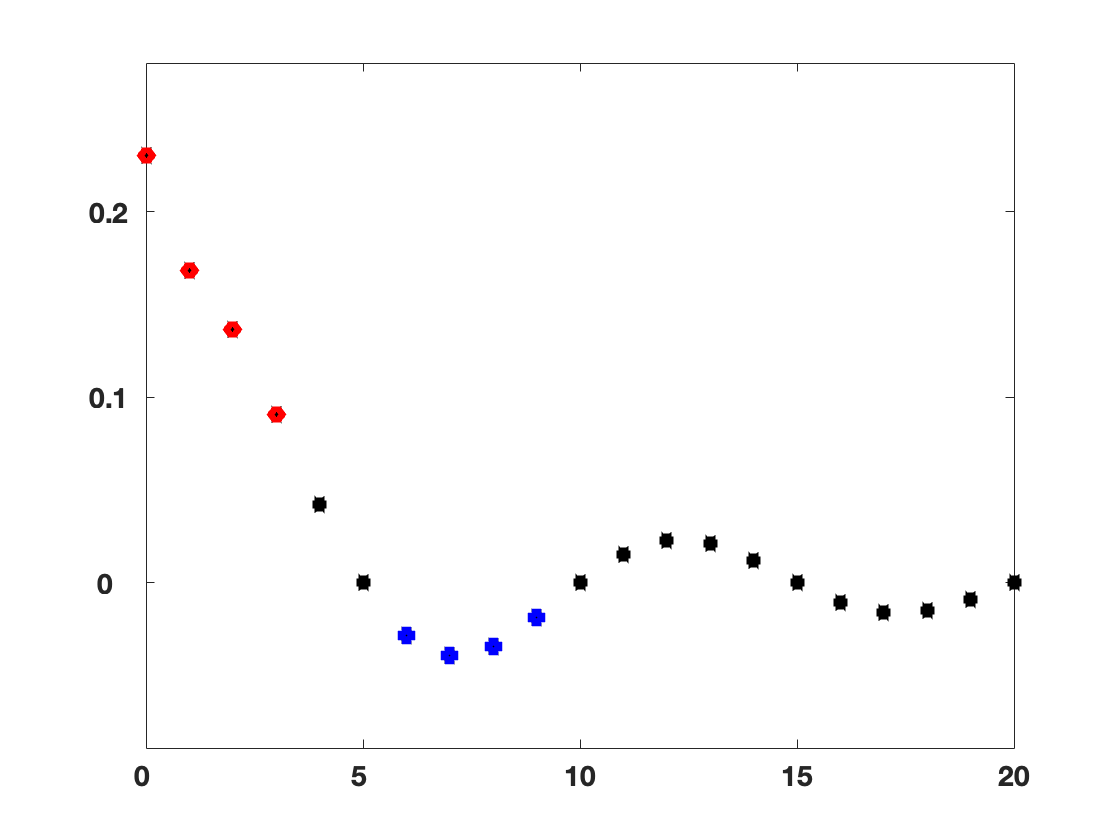}
\end{center}
\caption{Eigenvalues of $\bK$ for small-world graph with $p=0.05$ and
  $r=0.1$. The first four smallest and largest eigenvalues
are plotted in blue and red respectively.
}
\label{f.eigenval}
\end{figure}

Translational invariance of \eqref{clim} (cf.~\eqref{symmetries}) results  in SO(2) symmetry for
the Fourier coefficients in \eqref{F-eqn}. Consequently, the sum in \eqref{F-eqn} simplifies
to $3 \hat u_k |\hat u_k|^2$ ($k_1=\pm k$). The bifurcation equation is then obtained obtained
from the equation for $k=p$ in \eqref{F-eqn}
$$
\gamma \mu_p A- A|A|^2+\dots=0, \; A:=\hat u_p.
$$
From here we find $|A|\approx\sqrt{ \gamma\mu_p}$ and the solution bifurcating
from $0$ at $\beta=\mu_p^{-1}$ to leading order and up to a shift is given by
\begin{equation}\label{harmonic}
u_\gamma(x)\approx \sqrt{ \gamma\mu_p}\cos\left(2\pi px\right).
\end{equation}
The case of the simple eigenvalue $\mu_0$ is analyzed similarly. The main distinction is
that the bifurcating solution is constant in contrast to harmonics in
\eqref{harmonic} for
$k\neq 0$ (see Fig.~\ref{f.bif-sw}).

The eigenvalues of $\bK$ are known (cf.~\cite{MedPel2024}):
\begin{equation*}
\mu_k=\left\{ \begin{array}{ll} 2r(1-2p)+p, & k=0,\\
                (\pi k)^{-1}(1-2p) \sin\left(2\pi k r\right), & k\in \N.
\end{array}
                                                                \right.
                                                              \end{equation*}
  Fig.~\ref{f.eigenval} shows a few first $\mu_k$'s, where the
  four largest $(\mu_0 > \mu_1 > \mu_2 > \mu_3)$ and four smallest
  $(\mu_8 < \mu_9 < \mu_7 < \mu_{10})$
  eigenvalues are highlighted. These eigenvalues determine the bifurcating branches
  shown in Fig.~\ref{f.bif-sw}.
  The ordering of these eigenvalues explains the sequence, in which harmonics emerge in the
  AFM regime:
  the first bifurcating solution is $\cos(16\pi x)$, followed by $\cos(18\pi x)$, $\cos(14\pi x)$, and
  $\cos(20\pi x)$.
\begin{figure}[h]
    \centering
    \includegraphics[width=\linewidth]{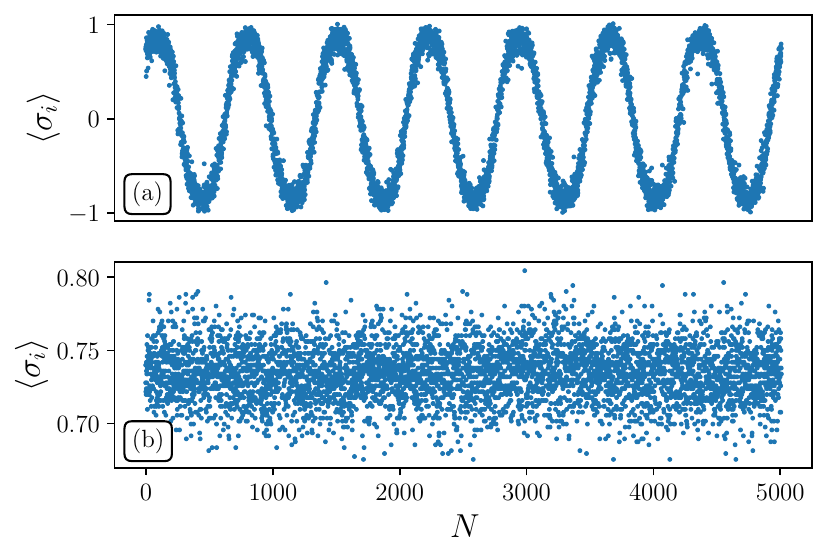}
    \caption{Spin configurations for: (a) AFM case ($J=-1$) slightly below $T=J\mu_8$, (b) FM case ($J=+1$) slightly below $T=J\mu_0$}
    \label{fig:stable-states}
  \end{figure}
  \begin{figure}[h]
    \centering
    \includegraphics[width=\linewidth]{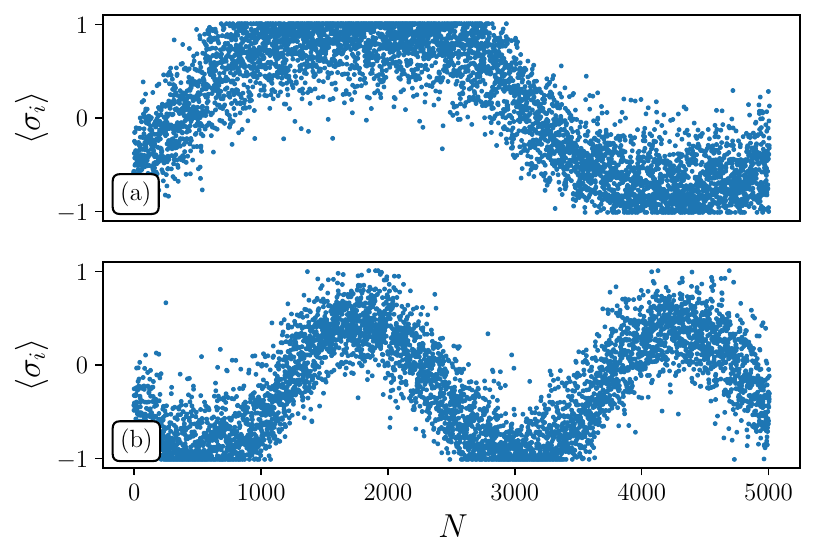}
    \caption{Spin configurations for: (a) FM case slightly below $T=J\mu_1$, (b) FM case slightly below $T=J\mu_2$}
    \label{fig:transient-states}
\end{figure}

  We performed Monte Carlo simulations using the Metropolis–Hastings algorithm~\cite[Ch.~4]{landau2021guide}
with parameters $r = 0.1$, $p = 0.05$, and $N = 5000$ spins. Starting from a random (uniform) spin
configuration, the system relaxes to the ground state corresponding to the principal bifurcating branch
at $\beta = \mu_0$ in the FM regime and $\beta = \mu_8$ in the AFM regime (see Fig.~\ref{f.bif-sw}).
To identify metastable states associated with $\mu_k$, $k \notin {0,8}$, the system was initialized
with $\operatorname{sign}\left(\bm{v}_k(x)\right)$, where $\bm{v}_k$ is the corresponding harmonic
mode (see insets in Fig.~\ref{f.bif-sw}). For temperatures just above $T = J\mu_k$, the system remains
for an extended period near the
corresponding eigenstate before drifting away and eventually relaxing to the ground state
(Fig.~\ref{fig:transient-states}). These simulations suggest that the coexistence of multiple
solutions at low temperatures gives rise to metastable behavior. As $T \to 0$,
the number of coexisting solutions of Eq.~\eqref{clim} can become arbitrarily large.

In this Letter, we have shown that a simple Ising model with nonlocal nearest-neighbor interactions
exhibits coexisting steady states leading to metastability. We expect that the analysis presented here
will help the understanding of phase transitions in the broad range of applications of the Ising model.

\textit{Acknowledgements}--
AA and GSM thank Jaime Cisternas for suggesting \verb|BifurcationKit|
and for sharing his code.
AA acknowledges discussions with Dmitriy Lyubshin and Mauro Mariani on Monte Carlo simulations.
The research of AA is supported by Basic Research Program at the HSE University and by
the Foundation for the Advancement of Theoretical Physics and Mathematics ``BASIS''
(grant No. 25-1-4-2-1). The work of GSM was supported in
part by the NSF (DMS-2406941).
Numerical simulations were performed at the HSE University. 

AA and GSM designed the study. GSM analyzed the phase transitions. AA performed the
numerical bifurcation analysis and Monte Carlo simulations.
Both authors interpreted the results and reviewed the manuscript.

The authors declare they have no conflicts of interest.

\textit{Data availability} --The data that support the findings of
this article are openly available at  \url{https://github.com/acubed3/Ising-Graphons}.

\bibliographystyle{apsrev4-2}
%
\end{document}